\documentclass{ifacconf}

\usepackage{graphicx}      
\usepackage{natbib}        
\usepackage{amssymb}
\usepackage{amsmath}
\usepackage{calrsfs}
\usepackage{algorithm,algorithmicx}
\usepackage{algpseudocode}

\newcommand{\norm}[1]{\left\lVert#1\right\rVert}
\newcommand{\col}[1]{\mathrm{col}\left(#1\right)}
\newcommand{\tr}[1]{\mathrm{tr}\left(#1\right)}
\newcommand{\vect}[1]{\mathrm{vec}\left(#1\right)}
\newcommand{\rank}[1]{\mathrm{rank}\left(#1\right)}
\newcommand{\blkdiag}[1]{\mathrm{blkdiag}\left(#1\right)}
\newcommand{\dett}[1]{\mathrm{det}\left(#1\right)}
\newcommand{\ellip}[1]{\mathcal{E}_f\left(#1\right)}

\makeatletter
\def\title@fmt#1#2{%
  \@ifundefined{@runtitle}{\global\def\@runtitle{#1}}{}%
  \@articletypesize
  \leavevmode\vphantom{Aye!}%
  \@articletype
  \vskip12\p@
  \begingroup
    \parindent=0pt
    \leftskip=0pt
    \rightskip=0pt
    \parfillskip=0pt
    \noindent\hspace*{0.175\textwidth}%
    \parbox{0.65\textwidth}{\centering
      {\@titlesize #1\,\hbox{$^{#2}$}\par}%
    }\par
  \endgroup
  \vskip\@undertitleskip
}
\makeatother

\begin{document}
\begin{frontmatter}

\title{A Unified Bayesian Framework for Data-Driven Smoothing, Prediction, and Control\thanksref{footnoteinfo}} 

\thanks[footnoteinfo]{This work was supported by the Lower Saxony Ministry for Science and Culture within the program zukunft.niedersachsen.}
\thanks{© 2026 the authors. This work has been accepted to IFAC for publication under a Creative Commons Licence CC-BY-NC-ND.}

\author[First]{Mingzhou Yin} 
\author[Second]{Andrea Iannelli} 
\author[First]{Seyed Ali Nazari} 
\author[First]{Matthias A. Müller}

\address[First]{Institute of Automatic Control, Leibniz University Hannover, 30167 Hannover, Germany (e-mail: yin@irt.uni-hannover.de, ali.nazari@stud.uni-hannover.de, mueller@irt.uni-hannover.de).}
\address[Second]{Institute for Systems Theory and Automatic Control, University of Stuttgart, 70569 Stuttgart, Germany (e-mail: andrea.iannelli@ist.uni-stuttgart.de)}

\begin{abstract}                
Extending data-driven algorithms based on Willems' fundamental lemma to stochastic data often requires empirical and customized workarounds. This work presents a unified Bayesian framework for linear systems that provides a systematic and general method for handling stochastic data-driven tasks, including smoothing, prediction, and control, via maximum a posteriori estimation. This framework formulates a unified trajectory estimation problem for the three tasks by specifying different types of trajectory knowledge. Then, a Bayesian problem is solved that optimally combines trajectory knowledge with a data-driven characterization of the trajectory from offline data for correlated input-output uncertainties with elliptical distributions. Under specific conditions, this problem is shown to generalize existing data-driven prediction and control algorithms. Numerical examples demonstrate the performance of the unified approach for all three tasks against other data-driven and system identification approaches.
\end{abstract}


\end{frontmatter}

\section{Introduction}

Data-driven analysis and control of linear systems, utilizing Willems' fundamental lemma \citep{Willems_2005}, have been a topic of considerable interest in recent years. When deterministic data are available, behavioral systems theory provides a powerful framework that allows for treating various data-driven problems in a unified manner, such as simulation and output matching \citep{Markovsky_2021, MARKOVSKY2022110008}. Since unique input-output behaviors can be identified with sufficiently informative data, equivalence can be shown for various data-driven formulations in the deterministic case \citep{Coulson_2019,Fiedler_2021}.

Approaches start to diverge when stochastic uncertainties arise. One path is to recover the low-dimensional structure in the deterministic case via projection \citep{Breschi_2023} or low-rank matrix approximation \citep{MARKOVSKY2022110008}. This is a practical choice, dating back to subspace identification \citep{van2012subspace}, but it requires knowledge of the model dimension and fails to encode specific stochastic properties for finite-sample optimality. Others apply specific techniques to particular data-driven tasks, such as distributionally robust optimization \citep{Coulson_2019_reg}, empirical regularizers \citep{dorfler2022bridging,Coulson_2019}, polynomial chaos expansions \citep{9999463}, Kalman filtering \citep{10886852,YIN202479}, kernel methods with truncated ARX models \citep{Chiuso_2025}, and maximum likelihood estimation \citep{yin2020maximum}. Yet, these techniques lack a coherent framework for general data-driven tasks with stochastic data.

This work introduces a unified Bayesian framework for linear systems that addresses stochastic data-driven problems through maximum a posteriori (MAP) estimation. This framework can be seen as an extension of the behavioral systems theory to stochastic data. A wide class of stochastic uncertainties is considered, which allows for input errors, correlated noise, and elliptical distributions. Specifically, when a stochastic signal matrix is available, the following three data-driven tasks are considered.
\begin{description}
    \item[(T1)]Smoothing: denoising measured input-output trajectories.
    \item[(T2)]Prediction: estimating future outputs given the immediate past input-output trajectory and future inputs.
    \item[(T3)]Optimal control: designing future inputs and outputs given the immediate past input-output trajectory and input-output references.
\end{description}

All these tasks can be posed as a trajectory estimation problem by specifying different types of observations and design requirements. A MAP estimator is then proposed that optimally combines the trajectory knowledge with a linear combination of offline data, following the characterization from Willems' fundamental lemma. The linear combination vector is treated as a hyperparameter and estimated using the maximum marginal likelihood method. An iterative solution to the hyperparameter estimation problem is derived using sequential quadratic programming (SQP). This demonstrates connections to existing algorithms in data-driven prediction and control under special conditions. Numerical examples show that the proposed framework achieves competitive performance against existing data-driven algorithms and classical approaches based on the system identification paradigm for all three tasks.

\textit{Notation.} For a sequence of matrices $X_1,\dots,X_n$, we denote the row-wise and diagonal-wise concatenation by $\col{X_1,\dots,X_n}$ and $\blkdiag{X_1,\dots,X_n}$, respectively. The rank, trace, determinant, log-determinant, and vectorization of a matrix are indicated by $\rank{\cdot}$, $\tr{\cdot}$, $\dett{\cdot}$, $\log\dett{\cdot}$, and $\vect{\cdot}$, respectively. For a vector $x$, $\norm{x}_P$ denotes the weighted $l_2$-norm $(x^\top Px)^{\frac{1}{2}}$. The symbols $\mathbb{N}_{[m,n]}$, $\mathbb{S}^n_+$, and $\mathbb{S}^n_{++}$ represent the set of natural numbers between $m$ and $n$, positive semidefinite, and positive definite matrices of dimension $n\times n$, respectively. The Kronecker product is indicated by $\otimes$.

\section{Background and Problem Formulation}

Consider a discrete-time linear time-invariant (LTI) system $y_t^0=Gu_t^0$ with $n_x$ underlying states, where $u_t^0\in\mathbb{R}^{n_u}$ and $y_t^0\in\mathbb{R}^{n_y}$ are the true inputs and outputs, respectively. Suppose the inputs and outputs are subject to uncertainties with $u_t=u_t^0+w_t$ and $y_t=y_t^0+v_t$, where $w_t$ and $v_t$ denote the actuation and measurement uncertainties, respectively. All the perturbed signals at time $t$ are denoted by $z_t:=\col{u_t,y_t}=z_t^0+\nu_t\in\mathbb{R}^n$, $n:=n_u+n_y$, where $z_t^0:=\col{u_t^0,y_t^0}$ is the unknown true signals and $\nu_t:=\col{w_t,v_t}$ denotes the uncertainties.

This paper considers the problem of obtaining the optimal estimate of a length-$L$ signal trajectory from possibly partial information of the corrupted trajectory and historical data of signal trajectories. In particular, assume that we have obtained $M$ full signal trajectories of length $L$: $\mathbf{z}^d_i := \col{z^d_{t_i+1},z^d_{t_i+2}\dots,z^d_{t_i+L}}\in\mathbb{R}^{nL}$, for $i=1,2,\dots,M$, where the superscript $d$ denoted historical data. Concatenating these trajectories column-wise formulates the signal matrix: $H:= \begin{bmatrix}\mathbf{z}^d_1&\mathbf{z}^d_2&\cdots&\mathbf{z}^d_M\end{bmatrix}\in\mathbb{R}^{nL\times M}$, which is referred to as the offline data in what follows. For analysis purposes, we also define the true signal matrix $H^0$ and the uncertainty matrix $\Delta_H$: $H=H^0+\Delta_H$.

The objective is to estimate another signal trajectory $\mathbf{z}^0 := \col{z_1^0,z_2^0\dots,z_L^0}\in\mathbb{R}^{nL}$ from offline data (instead of the model $G$) and existing knowledge of the trajectory. This trajectory is referred to as the online trajectory. The knowledge is expressed by
\begin{equation}
    \boldsymbol{\zeta} = \Phi\mathbf{z}^0 + \boldsymbol{\epsilon},
    \label{eq:maineq}
\end{equation}
where $\boldsymbol{\zeta}\in\mathbb{R}^p$ encodes direct observations and possibly design requirements of the trajectory $\mathbf{z}^0$, $\boldsymbol{\epsilon}\in\mathbb{R}^p$ represents the uncertainty of the knowledge, and $\Phi\in\mathbb{R}^{p\times nL}$ is a known transformation matrix.

By specifying different types of existing knowledge, \eqref{eq:maineq} encompasses a broad range of data-driven tasks in systems and control as follows.

\textbf{(T1)} Smoothing. The complete trajectory is available, but it is corrupted. In this case, we have $p=nL$,
\begin{align*}
    \Phi=\mathbb{I}_{nL},\ \boldsymbol{\zeta}=\col{z_1,z_2,\dots,z_L},\ \boldsymbol{\epsilon}=\col{\nu_1,\nu_2,\dots,\nu_L}.
\end{align*}

\textbf{(T2)} Prediction. The complete input sequence and the initial output sequence of length $L_0\geq l$ are available, where $l$ is the lag of $G$. Let $L':=L-L_0$. In this case, we have $p=nL_0+n_uL'$,
\begin{align*}
    \Phi&=\blkdiag{\mathbb{I}_{nL_0},\mathbb{I}_{L'}\otimes\begin{bmatrix}
    \mathbb{I}_{n_u}&\mathbf{0}_{n_u\times n_y}
\end{bmatrix}},\\
    \boldsymbol{\zeta}&=\col{z_1  ,\dots,z_{L_0},u_{L_0+1},\dots,u_L},\\
    \boldsymbol{\epsilon}&=\textrm{col}\,(\nu_1,\dots,\nu_{L_0},w_{L_0+1},\newline\dots,w_L).
\end{align*}
When $L_0\geq l$, from (14) of \cite{Iannelli_2021}, there exists $K$ such that $H^0=K\Phi H^0$. Noting that $\rank{AB}\leq\rank{B}$, this guarantees that $\rank{H^0}=\rank{\Phi H^0}$ in Assumption~\ref{ass:2} below.

\textbf{(T3)} Optimal control. An initial input-output sequence of length $L_0\geq l$ is available. We would like to design an input-output sequence $(z_t^0)_{t=L_0+1}^L$ such that the quadratic control cost $\sum_{t=L_0+1}^L\left(\norm{u_t^0-u_t^\mathrm{ref}}_R^2+\norm{y_t^0-y_t^\mathrm{ref}}_Q^2\right)$ is minimized, where $u_t^\mathrm{ref}\in\mathbb{R}^{n_u}$, $y_t^\mathrm{ref}\in\mathbb{R}^{n_y}$ are the input and output references and $R\in\mathbb{S}_{++}^{n_u}$, $Q\in\mathbb{S}_{++}^{n_y}$ are the input and output cost matrices, respectively. Let $\zeta_t^\mathrm{ctr}:=\col{u_t^\mathrm{ref},y_t^\mathrm{ref}}\in\mathbb{R}^n$. In this case, we have $p=nL$
\begin{align*}
    \Phi&=\mathbb{I}_{nL},\ \boldsymbol{\zeta}=\col{z_1,\dots,z_{L_0}  ,\zeta_{L_0+1}^\mathrm{ctr},\dots,\zeta_L^\mathrm{ctr}},\\
    \boldsymbol{\epsilon}&=\col{\nu_1,\dots,\nu_{L_0},\boldsymbol{\epsilon}^\mathrm{ctr}},
\end{align*}
where $\boldsymbol{\epsilon}^\mathrm{ctr}$ denotes the errors from the references.

Then, we are ready to state the main problem of this work.
\begin{prob}
    Given the offline data $H$ and the online trajectory observation $\boldsymbol{\zeta}$, obtain an optimal estimate $\hat{\mathbf{z}}$ of $\mathbf{z}^0$.
    \label{prob:1}
\end{prob}
\vspace{-1.5em}
The offline data $H$ will be incorporated into estimation as the model surrogate through a MAP framework, which will be detailed in Section~\ref{sec:map}.

\subsection{Elliptical Distributions and Noise Specification}

For tractability, elliptical distributions and stationary elliptical processes are considered in this work. An $m$-dimensional elliptical distribution, denoted by $\ellip{\mu,\Sigma}$, has a probability density function in the form of $p(x;\mu,\Sigma)=\dett\Sigma^{-\frac{1}{2}}f\left(\norm{x-\mu}_{\Sigma^{-1}}^2\right)$, where 
$f(\cdot)\in[0,+\infty)\to[0,+\infty)$ with $\lim_{x\to +\infty}f(x)=0$ and 
$\mu\in\mathbb{R}^m$ and $\Sigma\in\mathbb{S}_{++}^m$ are known as the location-scale parameters. Elliptical distributions are preserved under linear transformation, i.e., $x\sim\ellip{\mu,\Sigma}$ implies $Ax+b\sim\ellip{A\mu+b,A\Sigma A^\top}$ \citep{Hult_Lindskog_2002}. Such distributions include the multivariate normal distribution ($f(x)\propto\exp\left(-\frac{1}{2}x\right)$) and the multivariate Student's $t$-distribution ($f(x)\propto\left(1+x/\xi\right)^{-(\xi+m)/2}$ with parameter $\xi$). Stationary elliptical processes are defined as stationary processes with elliptical distributions \citep{bankestad2023variational}. 

Here, we slightly generalize the definition of elliptical distributions by allowing positive semidefinite $\Sigma$ to account for the case where the uncertainty is zero in some directions. When $\rank{\Sigma}=r<m$, consider the decomposition $\left(U_1,U_2,\Sigma_1\right)$ of $\Sigma$, where
\begin{equation}
    \Sigma=:\begin{bmatrix}
    U_1&U_2
\end{bmatrix}\begin{bmatrix}
    \Sigma_1&\mathbf{0}\\
    \mathbf{0}&\mathbf{0}
\end{bmatrix}\begin{bmatrix}
    U_1^\top\\U_2^\top
\end{bmatrix},\ 
\begin{bmatrix}
    U_1^\top\\U_2^\top
\end{bmatrix}\begin{bmatrix}
    U_1&U_2
\end{bmatrix}=\mathbb{I}_m,
\label{eq:semidef}
\end{equation}
$U_1\in\mathbb{R}^{m\times r}$, $U_2\in\mathbb{R}^{m\times (m-r)}$, and $\Sigma_1\in\mathbb{S}_{++}^r$, Then, the probability density function for singular $\Sigma$ is given by
\begin{equation*}
    p(x;\mu,\Sigma)=\begin{cases}
        \dett{\Sigma_1}^{-\frac{1}{2}}f\left(\norm{U_1^\top\boldsymbol{\delta}}_{\Sigma_1^{-1}}^2\right),&U_2^\top\boldsymbol{\delta}=\mathbf{0}_{m-r},\\
        0,&U_2^\top\boldsymbol{\delta}\neq\mathbf{0}_{m-r},
    \end{cases}
\end{equation*}
where $\boldsymbol{\delta}:=x-\mu$.

This work considers the following assumption.
\begin{assum}
     The random variable $\boldsymbol{\epsilon}$ is subject to zero-mean elliptical distributions: $\ellip{\mathbf{0}_p,\Sigma_\epsilon}$, where $\Sigma_\epsilon\in\mathbb{S}_+^p$. The random variables $\nu_t$ and $\nu_t^d$ are realizations of zero-mean stationary elliptical processes:
    \begin{align}
        \begin{bmatrix}
            \nu_t^d\\
            \nu_{t+\tau}^d\\
        \end{bmatrix}
        &\sim\ellip{\mathbf{0}_{2n},
        \begin{bmatrix}
            \Sigma_\nu^d(0)&\Sigma_\nu^d(\tau)\\
            \Sigma_\nu^d(\tau)&\Sigma_\nu^d(0)
        \end{bmatrix}
        },\\
        \begin{bmatrix}
            \nu_t\\
            \nu_{t+\tau}\\
        \end{bmatrix}
        &\sim\ellip{\mathbf{0}_{2n},
        \begin{bmatrix}
            \Sigma_\nu(0)&\Sigma_\nu(\tau)\\
            \Sigma_\nu(\tau)&\Sigma_\nu(0)
        \end{bmatrix}
        },
    \end{align}
    for $\tau\in\mathbb{N}$, where $\Sigma_\nu^d(\tau)=:\blkdiag{\Sigma_w^d(\tau),\Sigma_v^d(\tau)}\in\mathbb{S}_+^n$, $\Sigma_\nu(\tau)=:\blkdiag{\Sigma_w(\tau),\Sigma_v(\tau)}\in\mathbb{S}_+^n$, $\Sigma_w^d(\tau),\Sigma_w(\tau)\in\mathbb{S}_+^{n_u}$, and $\Sigma_v^d(\tau),\Sigma_v(\tau)\in\mathbb{S}_+^{n_y}$.
\end{assum}
We allow for different levels of uncertainty in the offline data and the online trajectory, as the system may operate under different conditions.
Depending on the data-driven tasks, we have $\Sigma_\epsilon=\Sigma_z$ for (T1) and $\Sigma_\epsilon=\Phi\Sigma_z\Phi^\top$ for (T2), where the $(i,j)$-th $n\times n$-block element of $\Sigma_z$ is $\Sigma_\nu(|i-j|)$. For
(T3), it is natural to encode the control objective as a design requirement that $z_t^0$ is subject to an elliptical distribution: $z_t^0\sim\ellip{\zeta_t^\mathrm{ctr},\Sigma^\mathrm{ctr}}$ for $t=L_0+1,\dots,L$, where $\Sigma^\mathrm{ctr}:=\blkdiag{R^{-1},Q^{-1}}\in\mathbb{S}_{++}^n$. Note that $R$ and $Q$ can be uniformly scaled to trade off against data uncertainties. In this case, we have $\Sigma_\epsilon=\blkdiag{\Gamma\Sigma_z\Gamma^\top,\mathbb{I}_{L'}\otimes\Sigma^\mathrm{ctr}}$, where $\Gamma=\begin{bmatrix}
    \mathbb{I}_{nL_0}&\mathbf{0}_{nL'}
\end{bmatrix}$.

\subsection{Willems' Fundamental Lemma and Deterministic Case}

We would like to estimate $\mathbf{z}^0$ by combining both direct knowledge of the trajectory through the observation $\boldsymbol{\zeta}$ and the behavior of the LTI system through the signal matrix $H$. To characterize $\mathbf{z}^0$ with the signal matrix $H$, we utilize Willems' fundamental lemma as the foundation.
\begin{thm}
\citep[Corollary 19]{IM-FD} Suppose $\rank{H^0}=n_uL+n_x$. The input-output sequence $\mathbf{z}^0$ is a trajectory of $G$ iff there exists $\mathbf{g}\in\mathbb{R}^M$, such that $\mathbf{z}^0=H^0\mathbf{g}$.
\label{thm:1}
\end{thm}
Theorem~\ref{thm:1} extends the original Willems' lemma \citep{Willems_2005} by allowing for non-Hankel $H$ and providing a necessary and sufficient rank condition. This theorem solves Problem~\ref{prob:1} in the deterministic case, i.e., when $\boldsymbol{\epsilon}=\mathbf{0}_p$ and $H=H^0$, which is given in the following corollary.
\begin{assum}
$\rank{H^0}=\rank{\Phi H^0}=n_uL+n_x$.
\label{ass:2}
\end{assum}
\begin{cor}
Suppose $\boldsymbol{\epsilon}=\mathbf{0}_p$, $H=H^0$, and Assumption~\ref{ass:2} is satisfied. The online trajectory is determined uniquely by $\mathbf{z}^0=H\mathbf{g}$, where $\mathbf{g}\in\mathbb{R}^M$ satisfies $\boldsymbol{\zeta} = \Phi H\mathbf{g}$.
\end{cor}
This corollary generalizes Proposition 9 in \cite{MARKOVSKY2022110008}. The proof remains valid with a linear transformation of $H$ instead of selected rows of $H$.

Assumption~\ref{ass:2} provides the identifiability condition for Problem~\ref{prob:1}. When uncertainties arise and $H^0$ is not available, the rank condition of $H^0$ can be verified by the sufficient condition in Theorem~2 of \cite{vanWaarde_2020}, i.e., $G$ is controllable and the inputs are collectively persistently exciting of order $(L+n_x)$. The rank condition of $\Phi H^0$ is satisfied for the problems considered in this work, as shown in the following subsection.

\section{The Maximum a Posteriori Framework}
\label{sec:map}

To solve Problem~\ref{prob:1}, this work considers a MAP framework, where $\boldsymbol{\zeta}$ is treated as the observation and the Willems' fundamental lemma characterization is used as prior information.

In detail, $\mathbf{z}^0$ is estimated by solving
\begin{equation}
    \hat{\mathbf{z}}=\mathrm{arg}\underset{\mathbf{z}^0}{\mathrm{max}}\ p\left(\mathbf{z}^0|\boldsymbol{\zeta}\right)=\mathrm{arg}\underset{\mathbf{z}^0}{\mathrm{max}}\ p\left(\boldsymbol{\zeta}|\mathbf{z}^0\right)p\left(\mathbf{z}^0\right).
    \label{eq:maporigin}
\end{equation}
The conditional distribution of $\boldsymbol{\zeta}$ given $\mathbf{z}^0$ is given by 
$$\boldsymbol{\zeta}|\mathbf{z}^0\sim\ellip{\Phi\mathbf{z}^0,\Sigma_\epsilon}.$$
The prior distribution of $\mathbf{z}^0$ can be obtained from Theorem~\ref{thm:1} using the equation $$\mathbf{z}^0=H\mathbf{g}-\Delta_H\mathbf{g},$$
where $\mathbf{g}$ is treated as a hyperparameter. To quantify the uncertainty of $\Delta_H\mathbf{g}$ given $\mathbf{g}$, we follow a similar derivation as in \cite{yin2020maximum} by vectorizing $\Delta_H$: 
$$\Delta_H\mathbf{g}=\left(\mathbf{g}^\top\otimes \mathbb{I}_{nL}\right)\vect{\Delta_H}.$$
Let $\vect{\Delta_H}\sim\ellip{\mathbf{0}_{nLM},\Sigma_d}$,
where $\Sigma_d\in\mathbb{S}_+^{nLM}$. For $i,j\in\mathbb{N}_{[1,ML]}$, define $i=:(i_1-1)L+i_2$, $j=:(j_1-1)L+j_2$, where  $i_1,j_1\in\mathbb{N}_{[1,M]}$ and $i_2,j_2\in\mathbb{N}_{[1,L]}$. Then, $\Sigma_d$ is given by
\begin{equation}
    \left(\Sigma_d\right)_{i,j}=\Sigma_\nu^d(|t_{i_1}+i_2-(t_{j_1}+j_2)|),
    \label{eq:sigd}
\end{equation}
where $\left(\Sigma_d\right)_{i,j}$ is the $(i,j)$-th $n\times n$-block element of $\Sigma_d$. 

This leads to $\Delta_H\mathbf{g}|\mathbf{g}\sim\ellip{\mathbf{0}_{nL},\Sigma_g(\mathbf{g})}$, where
\begin{equation}
    \Sigma_g(\mathbf{g}):=\left(\mathbf{g}^\top\otimes \mathbb{I}_{nL}\right)\Sigma_d\left(\mathbf{g}\otimes \mathbb{I}_{nL}\right)\in\mathbb{S}_+^{nL},
    \label{eq:sigg}
\end{equation}
and thus the prior distribution: $\mathbf{z}^0|\mathbf{g}\sim\ellip{H\mathbf{g},\Sigma_g(\mathbf{g})}$.

The following lemma presents an efficient approach to calculate $\Sigma_g(\mathbf{g})$. In particular, we discuss two common constructions of $H$, namely the Page construction \citep{DAMEN1982202} with $t_{i+1}=t_i+L$ and the Hankel construction \citep{van2012subspace} with $t_{i+1}=t_i+1$. Page construction has less noise correlation than the Hankel construction at the cost of fewer data columns in the signal matrix.
\begin{lem}
    Let $\left(\Sigma_g(\mathbf{g})\right)_{i,j}$ be the $(i,j)$-th $n\times n$-block element of $\Sigma_g(\mathbf{g})$. The prior scale parameter $\Sigma_g(\mathbf{g})$ is given by
    \begin{equation}
    \left(\Sigma_g(\mathbf{g})\right)_{i,j}=\sum_{k=1}^M\sum_{m=1}^M g_{k}g_{m}\Sigma_\nu^d(|t_k+i-(t_m+j)|).
    \label{eq:sigg4}
    \end{equation}
    For Page and Hankel constructions of $H$, \eqref{eq:sigg4} is equivalent to
    \begin{equation}
        \left(\Sigma_g(\mathbf{g})\right)_{i,j}=\sum_{\tau=1-M}^{M-1}\Sigma_\nu^d(|\tau l+i-j|)K(\tau)
    \label{eq:sigg2}
    \end{equation}
    with $l=L$ for the Page construction and $l=1$ for the Hankel construction,
    where $K(\tau):=\sum_{k=1}^{M-|\tau|}g_k g_{k+|\tau|}$, $\tau\in\mathbb{N}_{[1-M,M-1]}$ is the unnormalized empirical autocorrelation function of $\mathbf{g}$.
\end{lem}
\begin{pf}
Let $\mathbf{g}^\top\otimes \mathbb{I}_{nL}=:\begin{bmatrix}
    \gamma_1&\gamma_2&\cdots&\gamma_{LM}
\end{bmatrix}$, where $\gamma_i\in\mathbb{R}^{nL\times n}$. From \eqref{eq:sigg}, we have
\begin{equation*}
\Sigma_g(\mathbf{g})=\sum_{i=1}^{LM}\sum_{j=1}^{LM}\gamma_i\left(\Sigma_d\right)_{i,j}\gamma_j^\top.
\end{equation*}
According to the structure of $\left(g^\top\otimes \mathbb{I}\right)$, we have $\gamma_i=\mathbf{e}_{i_2}\otimes g_{i_1}$, $\gamma_j=\mathbf{e}_{j_2}\otimes g_{j_1}$, where $\mathbf{e}_i\in \mathbb{R}^L$ is the unit vector with $i$-th non-zero entry. This leads to
\begin{equation*}
\Sigma_g(\mathbf{g})=\sum_{i=1}^{LM}\sum_{j=1}^{LM}g_{i_1}g_{j_1}\left(\mathbf{e}_{i_2}\otimes\mathbb{I}_n\right)\left(\Sigma_d\right)_{i,j}\left(\mathbf{e}_{j_2}^\top\otimes \mathbb{I}_n\right).
\end{equation*}
From \eqref{eq:sigd}, we have
\begin{equation*}
\left(\Sigma_g(\mathbf{g})\right)_{i_2,j_2}=\sum_{i_1=1}^M\sum_{j_1=1}^M g_{i_1}g_{j_1}\Sigma_\nu^d(|t_{i_1}+i_2-(t_{j_1}+j_2)|),
\end{equation*}
which is equivalent to \eqref{eq:sigg4}. Consider variable transformations $\tau:=i_1-j_1$ and $k=\min\left(i_1,j_1\right)$. We note $g_{i_1}g_{j_1}=g_k g_{k+|\tau|}$, $t_{i_1}+i_2-(t_{j_1}+j_2)=\tau L+i-j$ for the Page construction, and $t_{i_1}+i_2-(t_{j_1}+j_2)=\tau+i-j$ for the Hankel construction. These directly lead to \eqref{eq:sigg2}. \qed
\end{pf}

Then, we are ready to state the MAP estimator. When both $\Sigma_\epsilon$ and $\Sigma_g(\mathbf{g})$ are positive definite, \eqref{eq:maporigin} becomes
\begin{equation}
    \hat{\mathbf{z}}(\mathbf{g})=\mathrm{arg}\underset{\mathbf{z}^0}{\mathrm{max}}\ f\left(\norm{\boldsymbol{\delta}_\epsilon}_{\Sigma_\epsilon^{-1}}^2\right)f\left(\norm{\boldsymbol{\delta}_g}_{\Sigma_g^{-1}(\mathbf{g})}^2\right),
    \label{eq:map1}
\end{equation}
where $\boldsymbol{\delta}_\epsilon=\boldsymbol{\zeta}-\Phi\mathbf{z}^0$ and $\boldsymbol{\delta}_g:=\mathbf{z}^0-H\mathbf{g}$. Recall that $f(\cdot)$ defines the probability density function of $\mathcal{E}_f$. The scaling factors $\dett\Sigma_\epsilon^{-\frac{1}{2}}$ and $\dett\Sigma_g^{-\frac{1}{2}}$ are omitted since they do not depend on $\mathbf{z}^0$.

When $\Sigma_\epsilon$ and/or $\Sigma_g(\mathbf{g})$ are singular, define $\left(U_{\epsilon,1},U_{\epsilon,2},\Sigma_{\epsilon,1}\right)$ and $\left(U_{g,1},U_{g,2},\Sigma_{g,1}(\mathbf{g})\right)$ by applying the decomposition \eqref{eq:semidef} on $\Sigma_\epsilon$ and $\Sigma_g(\mathbf{g})$, respectively. Then, the MAP estimator is given by the following constrained problem:
\begin{equation}
\begin{aligned}
    \hat{\mathbf{z}}(\mathbf{g})=\mathrm{arg}\underset{\mathbf{z}^0}{\mathrm{max}}&\ f\left(\norm{U_{\epsilon,1}^\top\boldsymbol{\delta}_\epsilon}_{\Sigma_{\epsilon,1}^{-1}}^2\right)f\left(\norm{U_{g,1}^\top\boldsymbol{\delta}_g}_{\Sigma_{g,1}^{-1}(\mathbf{g})}^2\right)\\
    \mathrm{s.t.}&\ U_{\epsilon,2}^\top\boldsymbol{\delta}_\epsilon=\mathbf{0},\ U_{g,2}^\top\boldsymbol{\delta}_g=\mathbf{0}.
    \label{eq:map2}
\end{aligned}
\end{equation}

When the uncertainties are Gaussian, i.e., $f(x)\propto\exp\left(-\frac{1}{2}x\right)$, \eqref{eq:map1} is equivalent to a quadratic program:
\begin{equation}
\hat{\mathbf{z}}(\mathbf{g})=\mathrm{arg}\underset{\mathbf{z}^0}{\mathrm{min}}\ \norm{\boldsymbol{\delta}_\epsilon}_{\Sigma_\epsilon^{-1}}^2 + \norm{\boldsymbol{\delta}_g}_{\Sigma_g^{-1}(\mathbf{g})}^2
\end{equation}
with the closed-form solution:
\begin{equation}
    \hat{\mathbf{z}}(\mathbf{g})=\left(\Phi^\top\Sigma_\epsilon^{-1}\Phi+\Sigma_g^{-1}(\mathbf{g})\right)^{-1}\left(\Phi^\top\Sigma_\epsilon^{-1}\boldsymbol{\zeta}+\Sigma_g^{-1}(\mathbf{g})H\mathbf{g}\right).
    \label{eqn:clsol1}
\end{equation}
Following a similar derivation as in \cite{yin2020maximum}, \eqref{eq:map2} also admits a closed-form solution under Gaussian uncertainties:
\begin{multline}
    \hat{\mathbf{z}}(\mathbf{g}) = \left(F^{-1}-F^{-1}V^\mathsf{T}(V F^{-1}V^\mathsf{T})^{-1}VF^{-1}\right)\mathbf{m}\\+F^{-1}V^\mathsf{T}(V F^{-1}V^\mathsf{T})^{-1}\mathbf{c},
    \label{eqn:clsol2}
\end{multline}
where
\begin{align*}
F&=\Phi^\top U_{\epsilon,1}\Sigma_{\epsilon,1}^{-1}U_{\epsilon,1}^\top\Phi+U_{g,1}\Sigma_{g,1}^{-1}U_{g,1}^\top,\\
\mathbf{m}&=\Phi^\top U_{\epsilon,1}\Sigma_{\epsilon,1}^{-1}U_{\epsilon,1}^\top\boldsymbol{\zeta}+U_{g,1}\Sigma_{g,1}^{-1}U_{g,1}^\top H\mathbf{g},\\
V&=\col{U_{\epsilon,2}^\top\Phi,U_{g,2}^\top},\\
\mathbf{c}&=\col{U_{\epsilon,2}^\top\boldsymbol{\zeta},U_{g,2}^\top H\mathbf{g}}.
\end{align*}

\subsection{Hyperparameter Estimation}

To apply the MAP estimator \eqref{eq:map1} or \eqref{eq:map2}, the hyperparameter $\mathbf{g}\in\mathbb{R}^M$ needs to be estimated as well. In this regard, we adopt the maximum marginal likelihood method. Consider the equation $\boldsymbol{\zeta} = \Phi H\mathbf{g} - \Phi\Delta_H\mathbf{g} + \boldsymbol{\epsilon}$. The hyperparameter $\mathbf{g}$ is estimated by maximizing the marginal likelihood: $\hat{\mathbf{g}}=\mathrm{arg}\mathrm{max}_\mathbf{g}\ p\left(\boldsymbol{\zeta}|\mathbf{g}\right)$, where $\boldsymbol{\zeta}|\mathbf{g}\sim\ellip{\Phi H\mathbf{g},\Phi\Sigma_g(\mathbf{g})\Phi^\top+\Sigma_\epsilon}$.

Let $\Psi(\mathbf{g}):=\Phi\Sigma_g(\mathbf{g})\Phi^\top+\Sigma_\epsilon$ and $\boldsymbol{\delta}_\zeta(\mathbf{g})=\boldsymbol{\zeta}-\Phi H\mathbf{g}$. If there exists $\mathbf{g}$ such that $\Psi(\mathbf{g})$ is positive definite, the hyperparameter estimation problem is given by
\begin{equation}
    \hat{\mathbf{g}}=\mathrm{arg}\underset{\mathbf{g}}{\mathrm{max}}\ \dett{\Psi(\mathbf{g})}^{-\frac{1}{2}}f\left(\norm{\boldsymbol{\delta}_\zeta(\mathbf{g})}_{\Psi^{-1}(\mathbf{g})}^2\right).
    \label{eq:mml1}
\end{equation}
When $\Psi(\mathbf{g})$ is singular for all $\mathbf{g}$, similar to \eqref{eq:map2}, let $\left(U_{\Psi,1},U_{\Psi,2},\Psi_1(\mathbf{g})\right)$ be the decomposition of $\Psi(\mathbf{g})$ according to \eqref{eq:semidef}. We have
\begin{equation}
\begin{aligned}
    \hat{\mathbf{g}}=\mathrm{arg}\underset{\mathbf{g}}{\mathrm{max}}&\ \dett{\Psi_1(\mathbf{g})}^{-\frac{1}{2}}f\left(\norm{U_{\Psi,1}^\top\boldsymbol{\delta}_\zeta(\mathbf{g})}_{\Psi_1^{-1}(\mathbf{g})}^2\right)\\
    \mathrm{s.t.}&\ U_{\Psi,2}^\top\boldsymbol{\delta}_\zeta(\mathbf{g})=\mathbf{0}.
\end{aligned}
\label{eq:mml2}
\end{equation}

The estimated hyperparameter $\hat{\mathbf{g}}$ is used in \eqref{eq:map1} or \eqref{eq:map2} to obtain the MAP estimate. This assumes certainty equivalence without considering the hyperparameter estimation error, which is common for Bayesian methods with hyperparameters.
The algorithm is summarized in Algorithm~\ref{al:1}.
\begin{algorithm}[htb]
\caption{MAP estimator for data-driven smoothing, prediction, and control}
    \begin{algorithmic}[1]
        \State \textbf{Given:} $H$, $\boldsymbol{\zeta}$, $\Sigma_\nu^d(\tau)$, $\Sigma_\epsilon$, $\Phi$, $f(\cdot)$
        \State Solve \eqref{eq:mml1} or \eqref{eq:mml2} for $\hat{\mathbf{g}}$, where $\Sigma_g(\mathbf{g})$ is given by \eqref{eq:sigg4} or \eqref{eq:sigg2}.
        \State Solve \eqref{eq:map1} or \eqref{eq:map2} for $\hat{\mathbf{z}}(\hat{\mathbf{g}})$.
        \State \textbf{Output:} $\hat{\mathbf{z}}(\hat{\mathbf{g}})$
    \end{algorithmic}
\label{al:1}
\end{algorithm}

\subsection{An Iterative Solution}
\label{sec:iter}

Problems \eqref{eq:mml1} and \eqref{eq:mml2} are non-convex even for Gaussian uncertainties. This subsection derives an SQP algorithm to solve \eqref{eq:mml1} and \eqref{eq:mml2} under the following assumption.
\begin{assum}
1) The uncertainties are Gaussian, 2) the Page construction is used, and 3) $\Sigma_\nu^d(\tau)=\mathbf{0}_{n\times n}$ for all $\tau>0$.
\label{ass:3}
\end{assum}
Assumption \ref{ass:3}.1 specifies a particular $f(\cdot)$, whereas Assumptions \ref{ass:3}.2 and \ref{ass:3}.3 lead to a block diagonal $\Sigma_g(\mathbf{g})$. This iterative algorithm can serve as a computationally efficient approximate solution to the problem in practice, even when Assumption~\ref{ass:3} is not satisfied. It also demonstrates the relations between the MAP framework and existing algorithms for data-driven prediction and control.

Under Assumption~\ref{ass:3}, \eqref{eq:mml1} is equivalent to
\begin{equation*}
    \hat{\mathbf{g}}=\mathrm{arg}{\mathrm{min}}_\mathbf{g}\ J(\mathbf{g}),\ J(\mathbf{g}):=\log\dett{\Psi(\mathbf{g})}+\norm{\boldsymbol{\delta}_\zeta(\mathbf{g})}_{\Psi^{-1}(\mathbf{g})}^2,
\end{equation*}
where $\Psi(\mathbf{g})=\norm{\mathbf{g}}_2^2\Phi\left(\mathbb{I}_L\otimes\Sigma_\nu^d(0)\right)\Phi^\top+\Sigma_\epsilon$ from \eqref{eq:sigg2}. Let $\Lambda:=\Phi\left(\mathbb{I}_L\otimes\Sigma_\nu^d(0)\right)\Phi^\top$ and $\hat{\mathbf{g}}_k$ be the $k$-th iterate of the iterative algorithm. We approximate $J(\mathbf{g})$ locally at $\hat{\mathbf{g}}_k$ with a quadratic function as
\begin{align*}
    J(\mathbf{g})\approx J(\hat{\mathbf{g}}_k)&+\tr{\Lambda\Psi^{-1}(\hat{\mathbf{g}}_k)}\left(\norm{\mathbf{g}}_2^2-\norm{\hat{\mathbf{g}}_k}_2^2\right)\\
    &-\norm{\Psi^{-1}(\hat{\mathbf{g}}_k)\boldsymbol{\delta}_\zeta(\hat{\mathbf{g}}_k)}_\Lambda^2\left(\norm{\mathbf{g}}_2^2-\norm{\hat{\mathbf{g}}_k}_2^2\right)\\
    &+\left(\norm{\boldsymbol{\delta}_\zeta(\mathbf{g})}_{\Psi^{-1}(\hat{\mathbf{g}}_k)}^2-\norm{\boldsymbol{\delta}_\zeta(\hat{\mathbf{g}}_k)}_{\Psi^{-1}(\hat{\mathbf{g}}_k)}^2\right),
\end{align*}
where we make use of the results that
\begin{align*}
    &\tfrac{\partial}{\partial x}\log\dett{xA+B}=\tr{A(xA+B)^{-1}}\\
    &\tfrac{\partial}{\partial x}(xA+B)^{-1}=-(xA+B)^{-1}A(xA+B)^{-1}.
\end{align*}
This leads to the SQP algorithm:
\begin{equation}
    \hat{\mathbf{g}}_{k+1}=\mathrm{arg}\underset{\mathbf{g}}{\mathrm{min}}\ c(\hat{\mathbf{g}}_k)\norm{\mathbf{g}}_2^2+\norm{\boldsymbol{\delta}_\zeta(\mathbf{g})}_{\Psi^{-1}(\hat{\mathbf{g}}_k)}^2,
    \label{eq:mapsp}
\end{equation}
where $c(\hat{\mathbf{g}}_k):=\tr{\Lambda\Psi^{-1}(\hat{\mathbf{g}}_k)}-\norm{\Psi^{-1}(\hat{\mathbf{g}}_k)\boldsymbol{\delta}_\zeta(\hat{\mathbf{g}}_k)}_\Lambda^2$.

The SQP algorithm for \eqref{eq:mml2} can be obtained similarly:
\begin{equation}
    \begin{aligned}
        \hat{\mathbf{g}}_{k+1}=\mathrm{arg}\underset{\mathbf{g}}{\mathrm{min}}&\ \bar{c}(\hat{\mathbf{g}}_k)\norm{\mathbf{g}}_2^2+\norm{U_{\Psi,1}^\top\boldsymbol{\delta}_\zeta(\mathbf{g})}_{\Psi_1^{-1}(\hat{\mathbf{g}}_k)}^2\\
        \mathrm{s.t.}&\ U_{\Psi,2}^\top\boldsymbol{\delta}_\zeta(\mathbf{g})=\mathbf{0},
    \end{aligned}
    \label{eq:mapsp2}
\end{equation}
where $\bar{c}(\hat{\mathbf{g}}_k):=\tr{\bar{\Lambda}\Psi_1^{-1}(\hat{\mathbf{g}}_k)}-\norm{\Psi_1^{-1}(\hat{\mathbf{g}}_k)U_{\Psi,1}^\top\boldsymbol{\delta}_\zeta(\hat{\mathbf{g}}_k)}_{\bar{\Lambda}}^2$ and $\bar{\Lambda}=U_{\Psi,1}^\top\Lambda U_{\Psi,1}$.

Problem \eqref{eq:mapsp} can be non-convex when the Hessian $c(\hat{\mathbf{g}}_k)\mathbb{I}_M+H^\top\Phi^\top\Psi^{-1}(\hat{\mathbf{g}}_k)\Phi H$ is not positive semidefinite, as commonly observed in SQP. Standard techniques in sequential convex programming can be applied to address this issue, such as projecting the Hessian onto the space of positive semidefinite matrices and utilizing the trust region or cubic regularization method \citep{Nesterov2006}.

If closed-form solutions are desired, one can further approximate \eqref{eq:mapsp} or \eqref{eq:mapsp2} by only running one iteration from the pseudoinverse solution to $\boldsymbol{\zeta}=\Phi H\mathbf{g}$: 
$$\hat{\mathbf{g}}_{0}=\hat{\mathbf{g}}_\mathrm{pinv}=H^\top\Phi^\top\left(\Phi HH^\top\Phi^\top\right)^{-1}\boldsymbol{\zeta}.$$
In this case, $\boldsymbol{\delta}_\zeta(\hat{\mathbf{g}}_0)=\mathbf{0}_p$ and $c(\hat{\mathbf{g}}_k)$, $\bar{c}(\hat{\mathbf{g}}_k)$ are positive.\footnote{Note that for $A,B\in\mathbb{S}^{n}_{++}$, there exists $D$ such that $B=DD^\top$, and $\tr{AB}=\tr{ADD^\top}=\tr{D^\top AD}=\sum_{i=1}^n d_i^\top A d_i>0$, where $d_i$ is the $i$-th column of $D$.} So, \eqref{eq:mapsp} and \eqref{eq:mapsp2} are guaranteed to be convex. Similar to \eqref{eqn:clsol1} and \eqref{eqn:clsol2}, convex quadratic programs with only equality conditions admit closed-form solutions, which we omit here for space constraints. A similar strategy is shown to perform well numerically for the prediction problem \citep{pmlr-v144-yin21a}.

\subsection{Relations to Existing Algorithms}
 
The SQP algorithm is connected to existing algorithms in data-driven prediction and control under the following additional assumption.
\begin{assum}
1) $\Sigma_w(0)=\Sigma_w^d(0)=\mathbf{0}_{n_u\times n_u}$, 2) $\Sigma_\nu(\tau)=\mathbf{0}_{n\times n}$ for all $\tau>0$, 3) $\Sigma_v(0)=\sigma^2\mathbb{I}_{n_y}$, $\Sigma_v^d(0)=\sigma_d^2\mathbb{I}_{n_y}$, and for the control task, 4) $Q=q\mathbb{I}_{n_y}$ and $R=r\mathbb{I}_{n_y}$. 
\label{ass:4}
\end{assum}
Assumption~\ref{ass:4}.1 assumes no actuation uncertainties as in most existing algorithms, whereas Assumptions~\ref{ass:4}.2 to \ref{ass:4}.4 leads to a diagonal $\Psi(\mathbf{g})$ and thus scalar weights in the objective function.

For the prediction task (T2), select
\begin{align*}
    U_{\Psi,1}^\top&=\begin{bmatrix}
    \mathbb{I}_{L_0}\otimes\begin{bmatrix}
        \mathbf{0}_{n_y\times n_u}&\mathbb{I}_{n_y}
    \end{bmatrix}&\mathbf{0}_{n_yL_0\times n_uL'}
\end{bmatrix},\\U_{\Psi,2}^\top&=\blkdiag{\mathbb{I}_{L_0}\otimes\begin{bmatrix}
        \mathbb{I}_{n_u}&\mathbf{0}_{n_u\times n_y}
    \end{bmatrix},\mathbb{I}_{n_uL'}}.
\end{align*}
Then, $\bar{\Lambda}=\sigma_d^2\mathbb{I}_{n_yL_0}$ and $\Psi_1(\hat{\mathbf{g}}_k)=\norm{\hat{\mathbf{g}}_k}_2^2\bar{\Lambda}+\sigma^2\mathbb{I}_{n_yL_0}$.
Define $\mathbf{y}_p:=U_{\Psi,1}^\top\boldsymbol{\zeta}$, $H_{yp}:=U_{\Psi,1}^\top\Phi H$, $\mathbf{u}:=U_{\Psi,2}^\top\boldsymbol{\zeta}$, and $H_u:=U_{\Psi,2}^\top\Phi H$. The first iterate of \eqref{eq:mapsp2} is equivalent to
\begin{equation}
    \begin{aligned}
        \hat{\mathbf{g}}_1=\mathrm{arg}\underset{\mathbf{g}}{\mathrm{min}}&\ \lambda\norm{\mathbf{g}}_2^2+\norm{\mathbf{y}_p-H_{yp}\mathbf{g}}_2^2\\
        \mathrm{s.t.}&\ \mathbf{u}=H_u\mathbf{g},
    \end{aligned}
    \label{eq:mappred}
\end{equation}
where $\lambda:=n_yL_0\sigma_d^2$, when $\boldsymbol{\delta}_\zeta(\hat{\mathbf{g}}_0)=\mathbf{0}_p$. This problem coincides with the data-driven predictor presented in \cite{lian2021adaptive}, which is based on minimizing the Wasserstein distance between $\mathbf{y}_p$ and $H_{yp}\mathbf{g}$. Note that by choosing different $\lambda$, \eqref{eq:mappred} covers a wide range of data-driven predictors; see \cite{yin2021data} for an overview.

For the optimal control task (T3), select
\begin{align*}
    U_{\Psi,1}^\top&=\mathrm{blkdiag}\,\big(\mathbb{I}_{L_0}\otimes\begin{bmatrix}
        \mathbf{0}_{n_y\times n_u}&\mathbb{I}_{n_y}
    \end{bmatrix},\mathbb{I}_{L'}\otimes\begin{bmatrix}
        \mathbf{0}_{n_y\times n_u}&\mathbb{I}_{n_y}
    \end{bmatrix},\\
    &\qquad\qquad\qquad\qquad\qquad\qquad\qquad\,\mathbb{I}_{L'}\otimes\begin{bmatrix}
        \mathbb{I}_{n_u}&\mathbf{0}_{n_u\times n_y}
    \end{bmatrix}\big),\\
    U_{\Psi,2}^\top&=\begin{bmatrix}
    \mathbb{I}_{L_0}\otimes\begin{bmatrix}
        \mathbb{I}_{n_u}&\mathbf{0}_{n_u\times n_y}
    \end{bmatrix}&\mathbf{0}_{n_uL_0\times nL'}
    \end{bmatrix}.
\end{align*}
Then, $\bar{\Lambda}=\blkdiag{\sigma_d^2\mathbb{I}_{n_yL},\mathbf{0}_{n_uL'\times n_uL'}}$ and $\Psi_1(\hat{\mathbf{g}}_k)=\norm{\hat{\mathbf{g}}_k}_2^2\bar{\Lambda}+\blkdiag{\sigma^2\mathbb{I}_{n_yL_0},q^{-1}\mathbb{I}_{n_yL'},r^{-1}\mathbb{I}_{n_uL'}}$.

Define $\col{\mathbf{y}_p,\mathbf{y}^\mathrm{ref},\mathbf{u}^\mathrm{ref}}:=U_{\Psi,1}^\top\boldsymbol{\zeta}$, $\col{H_{yp},H_{yf},H_{uf}}:=U_{\Psi,1}^\top\Phi H$, $\mathbf{u}_p:=U_{\Psi,2}^\top\boldsymbol{\zeta}$, and $H_{up}:=U_{\Psi,2}^\top\Phi H$. The first iterate of \eqref{eq:mapsp2} is equivalent to
\begin{equation}
    \begin{aligned}
        \hat{\mathbf{g}}_1=\mathrm{arg}\underset{\mathbf{g}}{\mathrm{min}}&\ r\norm{\mathbf{u}^\mathrm{ref}-H_{uf}\mathbf{g}}_2^2+\lambda_1\norm{\mathbf{y}^\mathrm{ref}-H_{yf}\mathbf{g}}_2^2\\&\qquad\qquad\quad+\lambda_2\norm{\mathbf{y}_p-H_{yp}\mathbf{g}}_2^2+\lambda_3\norm{\mathbf{g}}_2^2\\
        \mathrm{s.t.}&\ \mathbf{u}_p=H_{up}\mathbf{g},
    \end{aligned}
    \label{eq:mapctl}
\end{equation}
where
\begin{align*}
    \lambda_1&:=\left(\sigma_d^2\norm{\hat{\mathbf{g}}_0}_2^2+1/q\right)^{-1},\ \lambda_2:=\left(\sigma_d^2\norm{\hat{\mathbf{g}}_0}_2^2+\sigma^2\right)^{-1},\\
    \lambda_3&:=n_y\sigma_d^2\left(L'\lambda_1+L_0\lambda_2\right),
\end{align*}
when $\boldsymbol{\delta}_\zeta(\hat{\mathbf{g}}_0)=\mathbf{0}_p$. This problem has the same form as regularized data-enabled predictive control (DeePC) schemes; see \cite{Coulson_2019,Berberich_2021} and others. Selecting the regularization parameters in regularized DeePC is a non-trivial problem and significantly affects the performance \citep{dorfler2022bridging}. In this regard, \eqref{eq:mapctl} provides a meaningful method for choosing the regularization parameters without trial and error.
\begin{rem}
If we further assume no data uncertainties, i.e., $\sigma^2=\sigma_d^2=0$, it is easy to see that \eqref{eq:mapctl} becomes the unregularized DeePC problem: $\mathrm{min}_\mathbf{g}\ r\norm{\mathbf{u}^\mathrm{ref}-H_{uf}\mathbf{g}}_2^2+q\norm{\mathbf{y}^\mathrm{ref}-H_{yf}\mathbf{g}}_2^2$, s.t. $\mathbf{u}_p=H_{up}\mathbf{g}$, $\mathbf{y}_p=H_{yp}\mathbf{g}$.
\end{rem}
Despite the relations shown above, the proposed unified framework differs from existing algorithms in the following ways. 1) Existing algorithms typically solely adopt the prior estimate $H\hat{\mathbf{g}}$ as the trajectory estimate, whereas an additional MAP step is performed in this work. 2) This work provides more general results compared to existing algorithms. Note that multiple conditions (Assumptions~\ref{ass:3} and \ref{ass:4}) are required to demonstrate resemblance to existing algorithms. 3) This work unifies data-driven prediction and direct data-driven predictive control under the same MAP formulation.

\section{Numerical Examples}

This section presents three numerical examples covering data-driven smoothing, prediction, and control.\footnote{The codes are available at https://doi.org/10.25835/5f7y68pm.} For each task, we compare the following methods: 1) \textit{Bayes}: the proposed Bayesian framework (Algorithm~\ref{al:1}) implemented with a nonlinear solver, 2) \textit{Approx}: the approximate convex solution by implementing one SQP iteration as described in Section~\ref{sec:iter}, 3) \textit{N4SID}: the system identification paradigm where the model is identified with subspace identification by \textsc{Matlab} command \texttt{n4sid}, and 4) \textit{Proj}: the framework presented in \cite{Breschi_2023} where $H_{yf}$ is first projected and then deterministic behavioral systems theory is applied \citep{MARKOVSKY2022110008}. For \textit{N4SID}, we assume that the true model order is known and apply a Kalman filter on the identified state-space model to obtain state estimates. For $\textit{Proj}$, we apply indirect data-driven control with the derived predictor.

For all tasks, 100 Monte Carlo simulations are conducted with the following parameters: $n_x=10$, $n_u=n_y=1$, $L=40$, $L_0=10$, and $M=61$. Signal matrices are built using the Hankel construction with 100 data points. Random systems are generated by \textsc{Matlab} command \texttt{drss} unless otherwise stated. The systems are normalized to have an $\mathcal{H}_2$-norm of 1. Non-convex optimization problems are solved by \textsc{Matlab} command \texttt{fmincon}; convex optimization problems are solved by \textsc{Mosek}.

\begin{figure}
    \centering
    \includegraphics[width=\columnwidth]{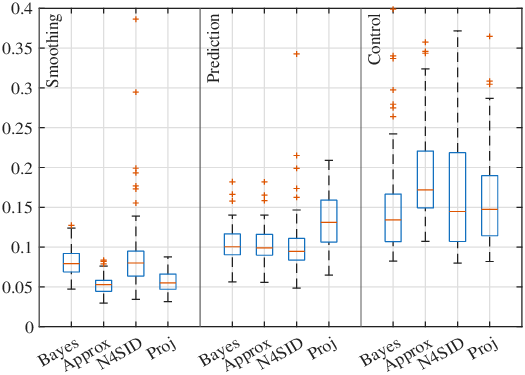}
    \caption{Performance comparison for data-driven smoothing, prediction, and control (lower is better). \textit{Bayes}: Algorithm~\ref{al:1}, \textit{Approx}: approximate solution with one iteration (Section~\ref{sec:iter}), \textit{N4SID}: system identification with N4SID, \textit{Proj}: two-step approach with projected signal matrices.}
    \label{fig:1}
\end{figure}

\textbf{(T1)} Smoothing. We assume that there are no actuation uncertainties. The measurement uncertainties are i.i.d. and subject to Student's $t$-distributions with $\xi=10$, $\Sigma_v^d(0)=10^{-4}$, and $\Sigma_v(0)=10^{-2}$. Figure~\ref{fig:1}(left) shows the boxplot of root mean squared errors for the smoothed outputs. Results show that \textit{Approx} and \textit{Proj} outperform \textit{Bayes} and \textit{N4SID}. The relatively poor performance of \textit{Bayes} may result from the local minimum issue with non-convex optimization.

\textbf{(T2)} Prediction. Both actuation and measurement uncertainties are i.i.d. Gaussian with $\Sigma_w^d(0)=\Sigma_v^d(0)=10^{-4}$ and $\Sigma_w(0)=\Sigma_v(0)=10^{-2}$. Figure~\ref{fig:1}(middle) shows the boxplot of root mean squared errors for the predicted outputs. The Bayesian framework performs better than \textit{Proj} and similar to \textit{N4SID}. The approximate solution yields nearly identical results to the non-convex solution.

\textbf{(T3)} Optimal control. We consider a discretized 1-D diffusion model specified as $x_{t+1}^1=(1-\beta)x_t^1+\alpha(x_t^2-x_t^1)+u_t$, $x_{t+1}^i=(1-\beta)x_t^i+\alpha(x_t^{i+1}+x_t^{i-1}-2x_t^i)$ for $i\in\mathbb{N}_{[2,9]}$, $x_{t+1}^{10}=(1-\beta)x_t^{10}+\alpha(x_t^9-x_t^{10})$, and $y_t = x_t^1$, where $x_t^i$ denotes the $i$-th component of $x_t$ and we select $\alpha=0.4$, $\beta=0.3$. We assume that there are no actuation uncertainties. The measurement uncertainties are Gaussian with $\Sigma_v^d(\tau)=\Sigma_v(\tau)=10^{-2}\times 0.95^\tau$. The control objective is set as $Q=5$, $R=0.5$, $\mathbf{y}^\mathrm{ref}=\col{\mathbf{1}_{10},-\mathbf{1}_{10},\mathbf{1}_{10}}$, and $\mathbf{u}^\mathrm{ref}=\mathbf{y}^\mathrm{ref}/d$, where $d$ denotes the DC gain of the system. Figure~\ref{fig:1}(right) shows the boxplot of the normalized root control cost for the true trajectory:
$$\textstyle J_\mathrm{ctr}=\sqrt{\frac{1}{L'}\sum_{t=L_0+1}^L\left(\frac{R}{Q}\norm{u_t^0-u_t^\mathrm{ref}}_2^2+\norm{y_t^0-y_t^\mathrm{ref}}_2^2\right)}.$$
Here, \textit{Bayes} achieves marginally better results than the other methods.

The computation times of the algorithms on an AMD Ryzen 7 8845H laptop are given in Table~\ref{tbl:1}. The high computation time for \textit{Bayes} in (T3) is due to the complex structure of $\Sigma_g(\mathbf{g})$ from correlated noise.
\begin{table}[ht]
    \renewcommand{\arraystretch}{1.1}
    \centering
    \caption{Median computation times}
    \vspace{-0.5em}
    \begin{tabular}{ccccc}
    \hline\hline
     & \textit{Bayes} & \textit{Approx} & \textit{N4SID} & \textit{Proj}\\\hline
    (T1) Smoothing & 0.89\,s & 0.31\,s & 0.03\,s & 0.22\,s \\
    (T2) Prediction & 0.68\,s & 0.24\,s & 0.04\,s & 0.21\,s \\
    (T3) Optimal control & 2.91\,s & 0.25\,s & 0.29\,s & 0.22\,s \\\hline\hline
    \end{tabular}
    \label{tbl:1}
\end{table}

Overall, the numerical examples illustrate that the proposed Bayesian framework achieves performance comparable to or slightly better than that of existing algorithms across all three tasks. It is worth noting that this is achieved through a unified direct data-driven formulation, whereas $\textit{Proj}$ relies on an indirect control formulation.

\section{Conclusion}

This work proposes solving the same Bayesian estimation problem for data-driven smoothing, prediction, and control with stochastic data. This approach first solves a hyperparameter estimation problem, followed by a maximum a posteriori step that combines the knowledge from offline data and the online trajectory. It unifies these three data-driven tasks and generalizes them to accommodate input errors, correlated uncertainties, and elliptical distributions. Further work may investigate uncertainty quantification, computational aspects, and applications to moving horizon estimation and receding horizon control.

\bibliography{ifacconf}

\begin{thebibliography}{25}
\providecommand{\natexlab}[1]{#1}
\providecommand{\url}[1]{\texttt{#1}}
\providecommand{\urlprefix}{URL }
\expandafter\ifx\csname urlstyle\endcsname\relax
  \providecommand{\doi}[1]{doi:\discretionary{}{}{}#1}\else
  \providecommand{\doi}{doi:\discretionary{}{}{}\begingroup
  \urlstyle{rm}\Url}\fi

\bibitem[{B{\r{a}}nkestad et~al.(2023)B{\r{a}}nkestad, Sj{\"o}lund, Taghia, and
  Sch{\"o}n}]{bankestad2023variational}
B{\r{a}}nkestad, M.M., Sj{\"o}lund, J., Taghia, J., and Sch{\"o}n, T.B. (2023).
\newblock Variational elliptical processes.
\newblock \emph{Transactions on Machine Learning Research}.

\bibitem[{Berberich et~al.(2021)Berberich, Köhler, Müller, and
  Allgöwer}]{Berberich_2021}
Berberich, J., Köhler, J., Müller, M.A., and Allgöwer, F. (2021).
\newblock Data-driven model predictive control with stability and robustness
  guarantees.
\newblock \emph{{IEEE} Transactions on Automatic Control}, 66(4), 1702--1717.

\bibitem[{Breschi et~al.(2023)Breschi, Chiuso, and Formentin}]{Breschi_2023}
Breschi, V., Chiuso, A., and Formentin, S. (2023).
\newblock Data-driven predictive control in a stochastic setting: a unified
  framework.
\newblock \emph{Automatica}, 152, 110961.

\bibitem[{Chiuso et~al.(2025)Chiuso, Fabris, Breschi, and
  Formentin}]{Chiuso_2025}
Chiuso, A., Fabris, M., Breschi, V., and Formentin, S. (2025).
\newblock Harnessing uncertainty for a separation principle in direct
  data-driven predictive control.
\newblock \emph{Automatica}, 173, 112070.

\bibitem[{{Coulson} et~al.(2019){Coulson}, {Lygeros}, and
  {Dörfler}}]{Coulson_2019}
{Coulson}, J., {Lygeros}, J., and {Dörfler}, F. (2019).
\newblock Data-enabled predictive control: In the shallows of the {DeePC}.
\newblock In \emph{European Control Conference ({ECC})}, 307--312.

\bibitem[{Coulson et~al.(2022)Coulson, Lygeros, and
  Dörfler}]{Coulson_2019_reg}
Coulson, J., Lygeros, J., and Dörfler, F. (2022).
\newblock Distributionally robust chance constrained data-enabled predictive
  control.
\newblock \emph{IEEE Transactions on Automatic Control}, 67(7), 3289–3304.

\bibitem[{Damen et~al.(1982)Damen, {Van den Hof}, and
  Hajdasinski}]{DAMEN1982202}
Damen, A., {Van den Hof}, P., and Hajdasinski, A. (1982).
\newblock Approximate realization based upon an alternative to the {Hankel}
  matrix: the {Page} matrix.
\newblock \emph{Systems \& Control Letters}, 2(4), 202--208.

\bibitem[{Dörfler et~al.(2023)Dörfler, Coulson, and
  Markovsky}]{dorfler2022bridging}
Dörfler, F., Coulson, J., and Markovsky, I. (2023).
\newblock Bridging direct \& indirect data-driven control formulations via
  regularizations and relaxations.
\newblock \emph{IEEE Transactions on Automatic Control}, 68(2), 883--897.

\bibitem[{Fiedler and Lucia(2021)}]{Fiedler_2021}
Fiedler, F. and Lucia, S. (2021).
\newblock On the relationship between data-enabled predictive control and
  subspace predictive control.
\newblock In \emph{European Control Conference ({ECC})}, 222--229.

\bibitem[{Hult and Lindskog(2002)}]{Hult_Lindskog_2002}
Hult, H. and Lindskog, F. (2002).
\newblock Multivariate extremes, aggregation and dependence in elliptical
  distributions.
\newblock \emph{Advances in Applied Probability}, 34(3), 587–608.

\bibitem[{Iannelli et~al.(2021)Iannelli, Yin, and Smith}]{Iannelli_2021}
Iannelli, A., Yin, M., and Smith, R.S. (2021).
\newblock Design of input for data-driven simulation with {Hankel} and {Page}
  matrices.
\newblock In \emph{IEEE Conference on Decision and Control (CDC)}, 139--145.

\bibitem[{Lian et~al.(2023)Lian, Shi, Koch, and Jones}]{lian2021adaptive}
Lian, Y., Shi, J., Koch, M., and Jones, C.N. (2023).
\newblock Adaptive robust data-driven building control via bilevel
  reformulation: An experimental result.
\newblock \emph{IEEE Transactions on Control Systems Technology}, 31(6),
  2420–2436.

\bibitem[{Markovsky and Dörfler(2021)}]{Markovsky_2021}
Markovsky, I. and Dörfler, F. (2021).
\newblock Behavioral systems theory in data-driven analysis, signal processing,
  and control.
\newblock \emph{Annual Reviews in Control}, 52, 42--64.

\bibitem[{Markovsky and Dörfler(2022)}]{MARKOVSKY2022110008}
Markovsky, I. and Dörfler, F. (2022).
\newblock Data-driven dynamic interpolation and approximation.
\newblock \emph{Automatica}, 135, 110008.

\bibitem[{Markovsky and Dörfler(2023)}]{IM-FD}
Markovsky, I. and Dörfler, F. (2023).
\newblock Identifiability in the behavioral setting.
\newblock \emph{IEEE Transactions on Automatic Control}, 68(3), 1667–1677.

\bibitem[{Nesterov and Polyak(2006)}]{Nesterov2006}
Nesterov, Y. and Polyak, B.T. (2006).
\newblock Cubic regularization of {Newton} method and its global performance.
\newblock \emph{Mathematical Programming}, 108(1), 177--205.

\bibitem[{Pan et~al.(2023)Pan, Ou, and Faulwasser}]{9999463}
Pan, G., Ou, R., and Faulwasser, T. (2023).
\newblock On a stochastic fundamental lemma and its use for data-driven optimal
  control.
\newblock \emph{IEEE Transactions on Automatic Control}, 68(10), 5922--5937.

\bibitem[{Smith et~al.(2024)Smith, Abdalmoaty, and Yin}]{10886852}
Smith, R.S., Abdalmoaty, M., and Yin, M. (2024).
\newblock Data-driven formulation of the {Kalman} filter and its application to
  predictive control.
\newblock In \emph{IEEE Conference on Decision and Control (CDC)}, 2633--2639.

\bibitem[{{Van Overschee} and {De Moor}(2012)}]{van2012subspace}
{Van Overschee}, P. and {De Moor}, B. (2012).
\newblock \emph{Subspace identification for linear systems: Theory,
  implementation, applications}.
\newblock Springer, New York, NY.

\bibitem[{{Van Waarde} et~al.(2020){Van Waarde}, {De Persis}, {Camlibel}, and
  {Tesi}}]{vanWaarde_2020}
{Van Waarde}, H.J., {De Persis}, C., {Camlibel}, M.K., and {Tesi}, P. (2020).
\newblock Willems’ fundamental lemma for state-space systems and its
  extension to multiple datasets.
\newblock \emph{IEEE Control Systems Letters}, 4(3), 602--607.

\bibitem[{Willems et~al.(2005)Willems, Rapisarda, Markovsky, and {De
  Moor}}]{Willems_2005}
Willems, J.C., Rapisarda, P., Markovsky, I., and {De Moor}, B.L.M. (2005).
\newblock A note on persistency of excitation.
\newblock \emph{Systems {\&} Control Letters}, 54(4), 325--329.

\bibitem[{Yin et~al.(2021)Yin, Iannelli, and Smith}]{pmlr-v144-yin21a}
Yin, M., Iannelli, A., and Smith, R.S. (2021).
\newblock Maximum likelihood signal matrix model for data-driven predictive
  control.
\newblock In \emph{Proceedings of the 3rd Conference on Learning for Dynamics
  and Control}, 1004--1014.

\bibitem[{Yin et~al.(2022)Yin, Iannelli, and Smith}]{yin2021data}
Yin, M., Iannelli, A., and Smith, R.S. (2022).
\newblock Data-driven prediction with stochastic data: Confidence regions and
  minimum mean-squared error estimates.
\newblock In \emph{European Control Conference (ECC)}, 853--858.

\bibitem[{Yin et~al.(2023)Yin, Iannelli, and Smith}]{yin2020maximum}
Yin, M., Iannelli, A., and Smith, R.S. (2023).
\newblock Maximum likelihood estimation in data-driven modeling and control.
\newblock \emph{IEEE Transactions on Automatic Control}, 68(1), 317--328.

\bibitem[{Yin et~al.(2024)Yin, Iannelli, and Smith}]{YIN202479}
Yin, M., Iannelli, A., and Smith, R.S. (2024).
\newblock Stochastic data-driven predictive control: Regularization,
  estimation, and constraint tightening.
\newblock \emph{IFAC-PapersOnLine}, 58(15), 79--84.

\end{thebibliography}

\appendix

\end{document}